# High-Density Horizontal Arrays of Single-Chirality Carbon Nanotubes

Yanzhao Liu[1]†, Zilong Qiu[1]†, Yuguang Chen[1]†, Nie Zhang[2], Bing Han[2], Huimin Yin[3], Bojun Liu[2], Min Lyu[1], Zhihong Li[4], Yiran Ma[1], Jian Sheng[1], Jiahui Shao[5], Zeyao Zhang[1,4], Li Ding[2], Hao Hong[5], Chuanhong Jin[3], Sheng Wang[2], Kaihui Liu[5], Xiaowei He[2], Lian-Mao Peng[2], Yan Li[1,4,6]*

**Affiliations:**

[1]Beijing National Laboratory for Molecular Science, College of Chemistry and Molecular Engineering, Peking University, Beijing 100871, China.

[2]Key Laboratory for the Physics and Chemistry of Nanodevices, School of Electronics, Peking University, Beijing 100871, China.

[3]School of Materials Science and Engineering, Zhejiang University, Hangzhou 310058, China.

[4]Institute of Carbon-Based Thin Film Electronics, Peking University, Shanxi, Taiyuan 030012, China.

[5]State Key Lab for Mesoscopic Physics, Frontiers Science Centre for Nano-optoelectronics, School of Physics, Peking University, Beijing 100871, China.

[6]Academy for Advanced Interdisciplinary Studies, Peking University, Beijing 100871, China.

*Corresponding author. Email: yanli@pku.edu.cn

†These authors contributed equally to this work.

**Abstract:** Highly ordered high-density arrays of single-chirality single-walled carbon nanotubes (SWCNTs) are greatly desired for exploring the intrinsic anisotropic properties and collective performance of such 1-dimensional (1D) nanomaterials. Here we present a Marangoni flow-induced self-assembly (MISA) strategy to fabricate monolayered SWCNT arrays achieving a packing density of ~200 $\mu m^{-1}$ and a 2-dimensional order parameter ($S_{2D}$) of ~ 0.95. Relying on its general compatibility with both organic and aqueous dispersions, we prepare single-chirality and enantiomer-pure SWCNT arrays from organic and aqueous dispersions resulting from the sorting processes. The anisotropic optical and electrical properties of the arrays are demonstrated by the polarization-dependent Rabi splitting as well as polarized near-infrared light emission and detection. With the great tolerance to solutions, substrates, and materials, as well as the feasibility and controllability, MISA shows great potential in the assembly of 1D nanomaterials.

One-dimensional (1D) materials exhibit strong structure anisotropy and consequently often possess distinctive quantum-confinement effects (*1, 2*). From the perspectives of both fundamental research and application, it's often essential to assemble them into ordered structures while retaining their intrinsic properties macroscopically (*3-5*). Such assemblies may also exhibit unique collective properties, e.g., single-walled carbon nanotube (SWCNT) arrays exhibit highly anisotropic optical behavior (*6-8*). Indeed, SWCNTs, featuring the atomically perfect cylindrical structures and 1D electronic states determined by their chiral index (*n,m*) (*9*), represent an ideal model system for the study of 1D nanomaterials' assembling.

Efforts for aligning SWCNTs have utilized external force fields (*10, 11*), delicate templates (*12, 13*) and interfaces (*14-18*). Tube density is an important criterion for many applications. For instance, high performance SWCNT-based integrated circuits require mono-layer SWCNT arrays with tube density >120 $\mu m^{-1}$ (*18-21*). For such purpose, methodology using interfacial technologies showed advantages. The dimension-limited self-alignment (DLSA) method confined SWCNTs at the two-dimensional organic-organic solvent interface (*18*), thus increasing the local concentration of SWCNTs at the interface and consequently resulting in wafer-size dense mono-layer arrays. Templated approaches such as DNA-directed assembly, produced highly aligned SWCNT arrays with a spacing of ~ 10 nm, yet it is feasible only on a very small scale (*12, 13*).

Recent progress in chirality-based sorting (*9, 19, 22*) motivates the assembly of single-chirality SWCNT arrays to explore the intrinsic structure dependent property and the emergence of new macroscopic properties from the structurally pure SWCNT assemblies (*23*). Furthermore, it provides a platform situated between isolated 1D materials and two-dimensional materials. Yet such assembly is still less studied. The controlled vacuum filtration method produced (6,5) SWCNTs films with the thickness of nanometers to sub-micrometers (*23*); however, it is difficult to readily prepare monolayered arrays and arrays with wafer-scale dimensions and high degree of alignment (*24*). Recently, a self-folding growth method yielded SWCNT arrays from a single tube with remarkably high density and high degree of alignment. However, the size scales are far from satisfactory (*25*). Currently, the high-purity chirality-specific SWCNTs are mainly achieved from sorting of aqueous dispersions (*9, 19, 22*) and more interestingly, enantiomer-based separations can also be realized via aqueous pathways (*26*). However, there are few reports on wafer-scaled high-density mono-layer SWCNT arrays obtained from aqueous SWCNT dispersions (*19*). Therefore, strategies compatible with aqueous systems are highly desirable.

Flow is a general phenomenon of liquid systems. Marangoni flow, which originates from the surface tension gradient, has shown its capability in the manipulation of liquid shapes (*27*), as well as in the assembly (*28*) and orientation (*29, 30*) of nanomaterials. With remarkable manipulating feasibility, Marangoni flow has great potential in SWCNT assembling. The unidirectional Marangoni flow presents a promising mechanism for driving SWCNTs toward the contact line. As SWCNTs accumulate at the meniscus, they tend to organize into liquid crystal domains. Combined with the one-dimensional confinement effect of the contact line, it may enable the fabrication of highly aligned, dense SWCNT arrays. We thus develop a Marangoni flow-induced self-assembly (MISA) method to prepare high-density, well-aligned SWCNT arrays. This strategy is applicable for both aqueous and organic SWCNT dispersions. By integrating chirality separation techniques, we demonstrate the fabrication of single-chirality and enantiomer-pure SWCNT arrays and investigate their advanced optical and optoelectronic properties emerging from the anisotropic and excitonic nature.

**Evaporation enabled Marangoni flow for SWCNT alignment**

Fig. 1A shows the general design of MISA method. The directional Marangoni flow is triggered by the differential evaporation of binary and miscible solvents. With the assistance of Marangoni flow toward the gas-solid-liquid three-phase contact line, SWCNTs accumulate and self-assemble at the meniscus, aligning horizontally along the contact line, and subsequently deposit onto the substrate. To direct the Marangoni flow toward the contact line and ensure its strength, we rationally choose the components in the binary solvents by two criteria: first, the less volatile component in the binary solvents has higher surface tension than the more volatile one; second, the surface tension and volatility difference between the two components are large enough. The less volatile component enriches near the contact line due to its lower evaporation rate. As a result, the liquid surface near the contact line has higher surface tension, and the Marangoni flow is formed toward the contact line. This Marangoni flow plays a key role in the self-assembly of the SWCNTs by pushing the SWCNTs to pre-organize at the contact line, form nematic liquid crystal assemblies, and eventually deposit on the substrate along the contact line. Under optimized binary solvent composition (see Supporting Information for experimental details), well-aligned SWCNTs are deposited as a continuous film at macroscopic horizontal dimensions of wafer scale. Its general alignment is evidenced by the cross-polarized optical microscopy images (Fig. S1). In contrast, if the solvent has only one component (e.g. pure toluene) or the less volatile component in the binary solvents presents lower surface tension than the more volatile component (e.g. toluene and n-decane), solutal Marangoni flow will be absent or the direction of Marangoni flow will be reversed, hence ordered assembly of SWCNTs will not be obtained (Fig. S2).

The MISA strategy is compatible with both organic and aqueous SWCNT dispersions. SWCNT arrays are self-assembled from single-stranded DNA (ssDNA)-wrapped SWCNTs dispersed in mixed water and methanol (~80:20 v/v) (Fig. 1B & Fig S3), or from conjugated polymer-wrapped SWCNTs dispersed in mixed toluene and nitrobenzene (Fig. 1C). The ssDNA or conjugated polymer can be removed by annealing in $H_2$/Ar atmosphere.

The cross-sectional transmission electron microscopy image reveals that the prepared aligned SWCNTs formed a monolayer (Fig. 1D), which is beneficial for electronic applications (*14, 18*). Statics of the atomic force microscopy (AFM) images (Fig. 1E) show that the density is ~ 100 - 200 /μm for the SWCNT arrays obtained from organic and aqueous dispersions, which meet the requirements of high performance integrated circuits (*20, 21*). The degree of alignment of SWCNT arrays is assessed by the two-dimensional order parameter $S_{2D}$ (*31*),

$$S_{2D} = \langle 2\cos^2\theta_d - 1 \rangle \tag{1}$$

where $\theta_d$ is the difference between the SWCNT's orientation angle and the overall SWCNTs' average orientation angle, with $S_{2D}$ = 1 corresponding to a perfect alignment and $S_{2D}$ = 0 corresponding to completely random orientation. By measuring the polarized Raman spectra with various polarization directions of the incident laser and fitting the G-band Raman intensity (Fig. 1F, S4, S5), the ratio of the maximum and minimum Raman intensity is calculated to be 40, and consequently the average deviation of the orientation angle is ±9° and the $S_{2D}$ is as high as 0.95. Compared with reported results (*19*), our MISA-prepared SWCNT arrays are among the best in terms of density and degree of alignment (Fig. 1G and S6).

It needs to be noticed that the conjugated polymer extraction can directly obtain high purity semiconducting SWCNTs (*32*). Then we fabricated field-effect transistors (FETs) from SWCNT arrays assembled from organic dispersions. The FET device shows an on-state current of 0.8 mA/μm, a peak transconductance of 1.4 mA/μm at $V_{ds}$ = -1 V, and a low-bias current on/off ratio over $10^4$ (Fig. S7).

The MISA method is compatible with various substrates, dispersants, and SWCNTs of different diameters (Fig. S8 & Fig. S9). It's valid for fabricating monolayered and multilayered arrays by controlling the concentrations and withdrawal speed (Fig. S10). By repeatedly assembling, well-controlled multilayers can also be produced. The resultant films can also be transferred to other substrates. By controlled transfer, the stacking of arrays at designed twist angles is fabricated (Fig. S11).

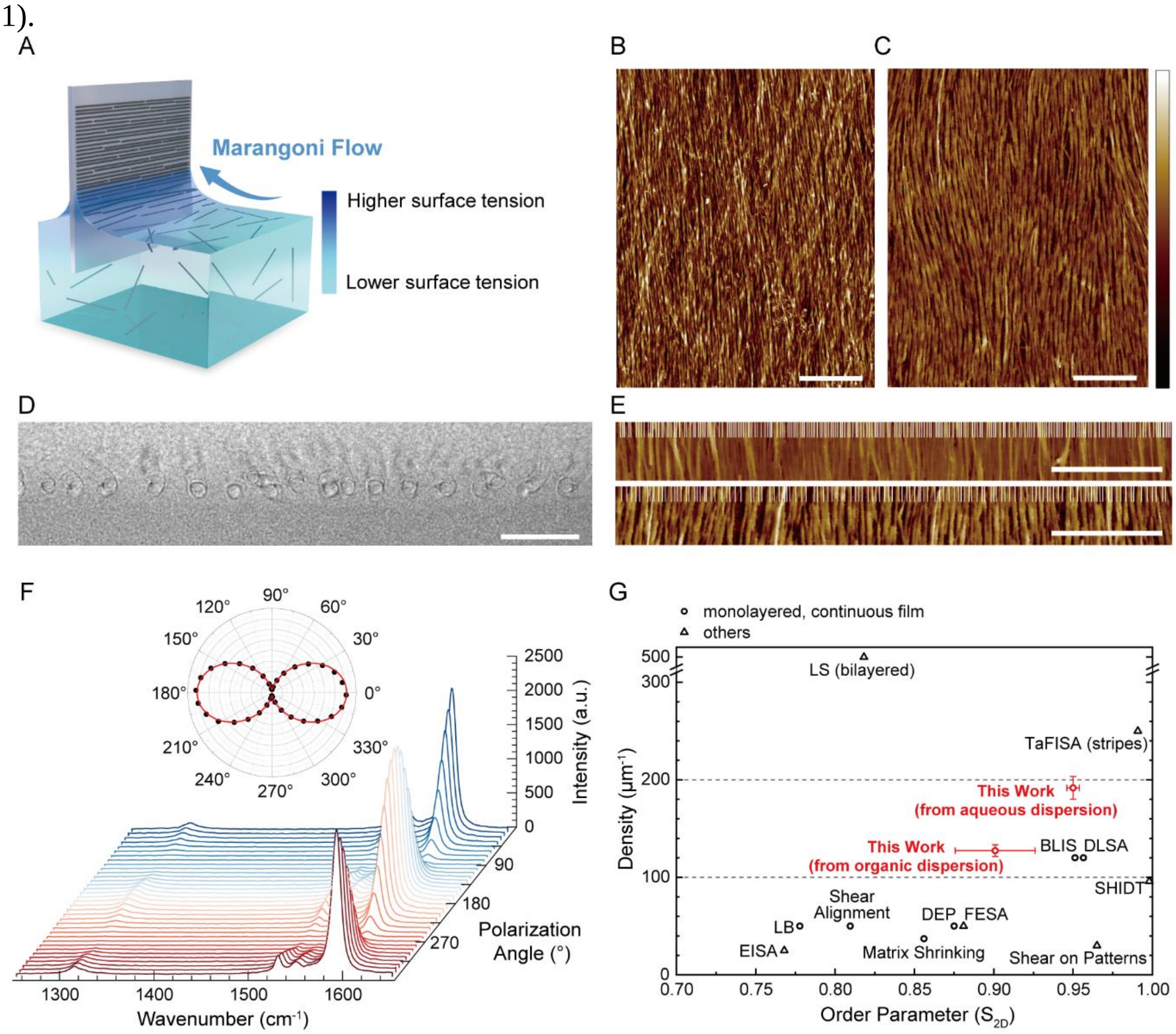

**Fig. 1. Marangoni flow-induced self-assembly (MISA) method for preparing high-density SWCNT arrays from both aqueous and organic dispersions.** (**A**) Scheme of the MISA process. (**B** and **C**) AFM images of high-density SWCNT arrays prepared from aqueous dispersions (B) and organic dispersions (C). Scale bars in (B) and (C) are 400 and 200 nm, respectively. Height range: -3 to 3 nm. (**D**) Cross-sectional TEM image of aligned SWCNTs obtained from organic dispersions. Scale bar: 10 nm. (**E**) AFM images for the density calculation of SWCNT arrays acquired from aqueous dispersions (top) and organic dispersions (bottom). Scale bar: 200 nm. (**F**) Polarized Raman spectra of aligned SWCNTs obtained from organic dispersions. Inset: Polar plot of the G band intensity for different laser polarizations. Incident laser: 532 nm. (**G**) Comparison of the packing density and $S_{2D}$ of the monolayered SWCNT arrays prepared by MISA with other methods (see data set and references in SI.).

**Preparation of single-chirality and enantiomer-pure SWCNT arrays with high density**

We prepare organic dispersions of (6,5), (9,8), and (7,5) enriched SWCNTs by conjugated polymer extraction from specific SWCNT powder samples (see details in SI) (*33*) (Fig. S12-S14). Then SWCNT arrays of the corresponding (*n,m*)-enriched SWCNTs are assembled and characterized (Fig. 2 A-H, Fig. S15-19). Let's take (6,5)-enriched array as an example. AFM characterization (Fig. 2A) confirms that the array is monolayered with a high packing density. Polarized Raman spectra demonstrate a high degree of alignment of the array (Fig. 2B and Fig. S15). The polar plot of the photoluminescence (PL) intensity of (6,5) SWCNT (Fig. 2C) and the corresponding polarized PL emission spectra (Fig. S16) indicate the anisotropy of the array. The 2-dimensional (2D) PL of (6,5) array is shown in Fig. S17. In the polarized absorption spectra (Fig. 2D), when the polarization of the incident light is parallel to the alignment direction, almost only the $E_{11}$ electronic transition of (6,5) and its phonon-assisted sideband peak of SWCNTs are observed, indicating the chirality purity of the array. Due to the high degree of alignment, the $E_{11}$ intensity significantly diminishes when the incident light is polarized perpendicular to the alignment direction.

Enantiomer-pure SWCNT arrays with a high density have yet to be realized. Such materials simultaneously exhibit dichroism and birefringence (both linear and circular), which may display unique properties (*34, 35*). We apply the MISA method to aqueous SWCNT dispersions to prepare enantiomeric SWCNT arrays. Utilizing a polyethylene glycol/salt-based aqueous two-phase extraction method (*26*), (−)(8,3), (−)(6,5), and (+)(6,5) SWCNT dispersions are obtained (Fig. S20-S22) and then assembled into arrays by the MISA method. AFM images clearly show the formation of aligned SWCNTs (Fig. 2I, 2M, and 2Q). The high degree of alignment is indicated by the Raman G band intensity dependence on the polarizations of the incident laser (Fig. 2J, 2N, and 2R) and the PL intensity dependence on the polarizations (Fig. 2K, 2O, and 2S) with the corresponding polarized spectra shown in Fig. S23-S25. The 2D PL spectra of the arrays are shown in Fig. S26-S28. The circular dichroism (CD) spectra (Fig. 2L, 2P, and 2T) reveal that the SWCNTs in the array are enantiomeric, and CD signals corresponding to the $E_{ii}$ (i = 2, 3, 4) transitions are observed. As expected, the CD signals of arrays composed of (−)(6,5) and (+)(6,5) SWCNTs have opposite signs.

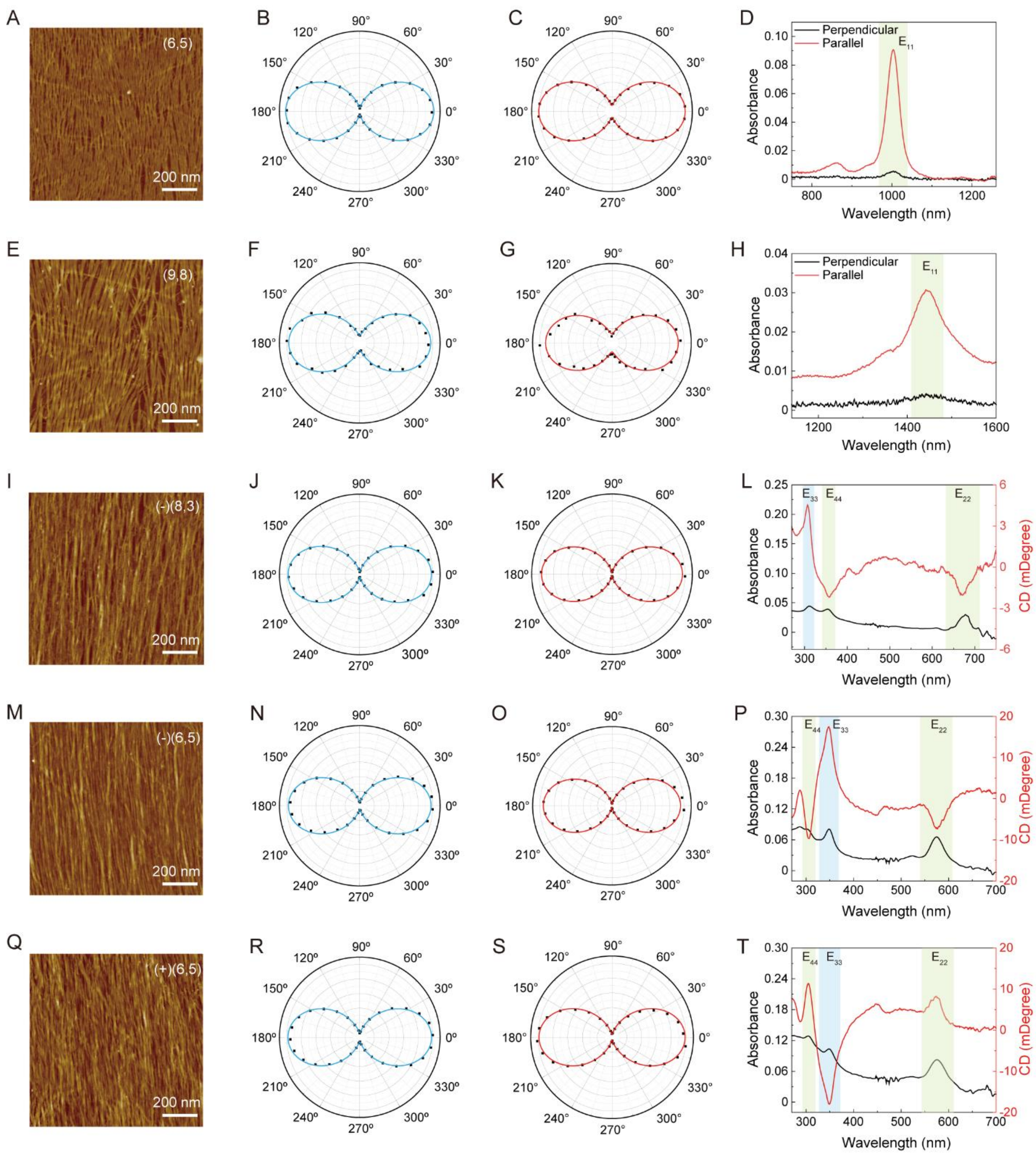

**Fig. 2. Preparation of single-chirality and enantiomer-pure SWCNT arrays with high density. (A** to **D)** (6,5) SWCNT array (prepared from organic dispersions): (A) AFM image; (B) Polar plot of Raman G band intensities; (C) Polar plot of photoluminescence intensities; (D) Polarized absorption spectra. **(E** to **H)** (9,8) SWCNT array (prepared from organic dispersions): (E) AFM image; (F) Polar plot of Raman G band intensities; (G) Polar plot of photoluminescence intensities; (H) Polarized absorption spectra. **(I** to **L)** (–)(8,3) SWCNT array (prepared from aqueous dispersions): (I) AFM image; (J) Polar plot of Raman G band intensities; (K) Polar plot of photoluminescence intensities; (L) CD & UV-Vis spectrum. **(M** to **P)** (–)(6,5) SWCNT array (prepared from aqueous dispersions): (M) AFM image; (N) Polar plot of Raman G band intensities;

(O) Polar plot of photoluminescence intensities; (P) CD & UV-Vis spectrum. **(Q** to **T)** (+)(6,5) SWCNT array (prepared from aqueous dispersions): (Q) AFM image; (R) Polar plot of Raman G band intensities; (S) Polar plot of photoluminescence intensities; (T) CD & UV-Vis spectrum.

### Anisotropic optical and optoelectronic properties of (6,5) SWCNT arrays

SWCNTs possess one-dimensional excitons with considerable oscillator strengths (*36*) and sharp excitonic transitions (*37*), providing a platform for studying strong coupling physics under extreme quantum confinement. We place (6,5) SWCNT arrays in an optical microcavity (Fig.3A) and illuminate the microcavity with polarized white light. When the incident light is polarized along the alignment direction, upper and lower branches are observed in the angle-resolved reflection spectra, indicating the formation of exciton-polaritons (Fig. 3C). The Rabi splitting energy, minimum energy discrepancy between the upper and lower branches, is about 100 meV (red line in Fig.3B), indicating that the coupling strength lies in the strong coupling regime. According to the reported relationship between the Rabi splitting energy and the array thickness, the Rabi splitting energy we observe is higher than the extrapolated results with the same thickness (*23*). Instead, when the array is illuminated with light polarized perpendicular to the alignment direction, no splitting appears in the angle-resolved reflectance spectrum (black line in Fig. 3B and Fig. 3D), indicating weak or no coupling between the array and the microcavity photons. These results highlight the anisotropic optical properties of SWCNT arrays and suggest that we can control the strong and weak coupling between light and SWCNTs by adjusting the polarization of the incident or reflected light.

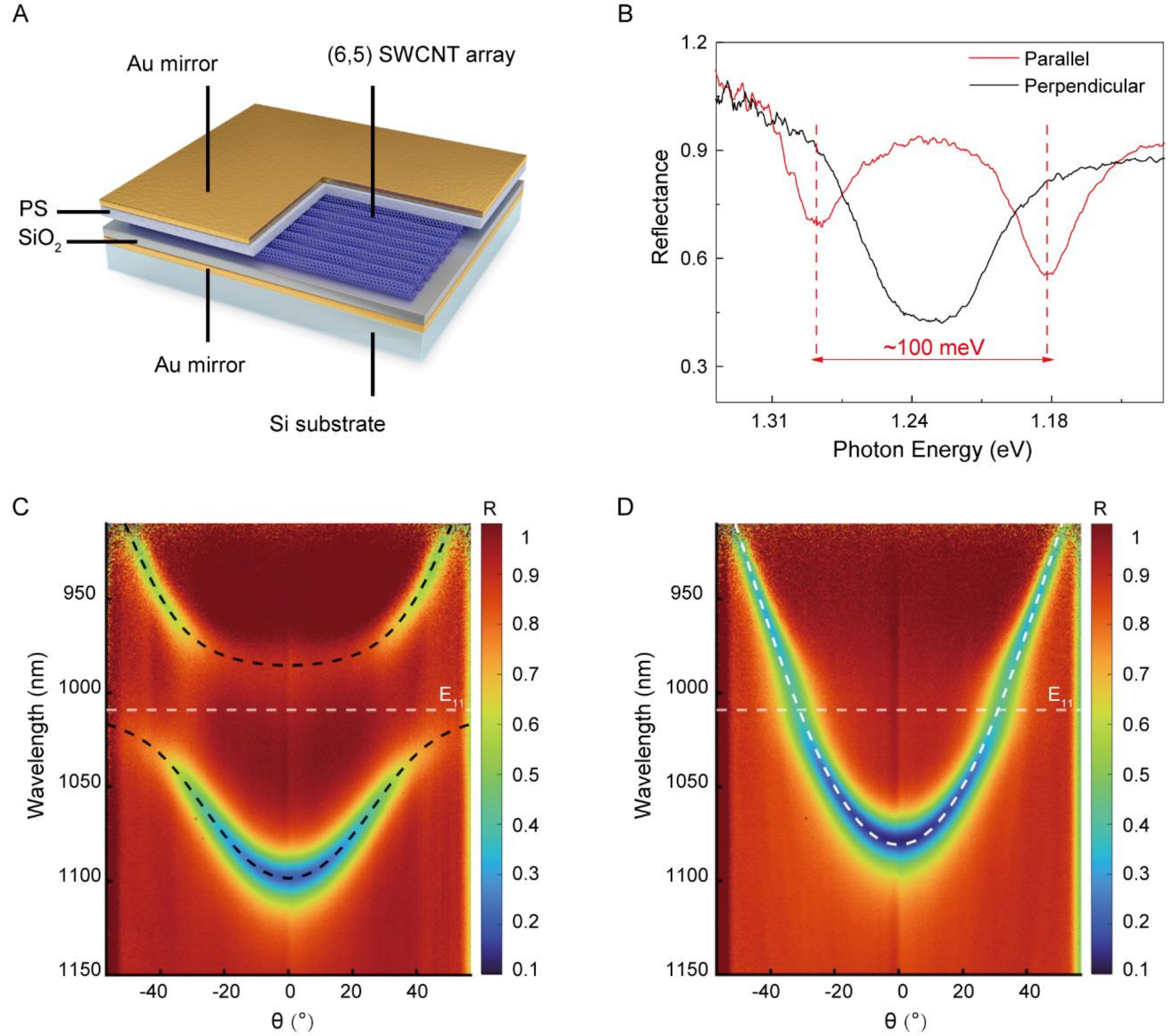

**Fig. 3. Polarization-dependent coupling between light and (6,5) SWCNT arrays in a microcavity**. (**A**) Scheme and structure of the microcavity containing (6,5) SWCNT arrays. (**B**) Reflected spectra of the cavity at zero detuning when the polarization of the incident light is parallel (red) and perpendicular (black) to the alignment direction. Splitting is observed when the incident light is polarized parallel to the alignment direction. (**C** and **D**) Angle-resolved reflection spectra of a microcavity in which a (6,5) SWCNT array is positioned when the incident light is polarized parallel (C) and perpendicular (D) to the alignment direction.

Using the (6,5) SWCNT arrays prepared from organic dispersions, we fabricate near-infrared (NIR) light detection diodes. Fig. 4A illustrates the scheme of the device. (6,5) SWCNT arrays are used as channel materials and are in contact with Pd and Hf electrodes, respectively. A layer of $HfO_2$ is deposited on the Hf electrode to improve the contact between Hf and aligned SWCNTs. Polarization-dependent detection of NIR light within a specific band is achieved by the high degree of order of the array (Fig. 4B). The incident light is set to be 1010 nm, which matches the $E_{11}$ transition of (6,5) SWCNTs. As the polarization direction of the incident laser is altered, the magnitude of the photocurrent changes significantly, resulting in a ratio of approximately 2.3 between the maximum and minimum photocurrents.

The near-infrared (NIR) band (1000-1600 nm) is significant for applications in communication, sensing, and imaging. Semiconducting SWCNTs are of interest as they display PL within this range. Their emission wavelength and quantum yield can be tuned through covalent sidewall modification, a strategy that has been extensively explored both for single-chirality SWCNTs in dispersion and for thin-film networks (*38, 39*). However, due to the disorder of SWCNTs in such samples, the PL generated by the defect sites is isotropic, losing the information about the polarization. Remarkably, we directly create $sp^3$ defects in (6,5) SWCNT arrays. When excited by a 575 nm laser, a new strong peak appears at 1270 nm, corresponding to the $E_{11}^{*-}$ transition resulting from defect sites (Fig. 4C). Moreover, both the PL corresponding to $E_{11}$ and $E_{11}^{*-}$ transitions are significantly polarized (Fig. 4C and 4D). The introduction of polarization information would open up more possibilities for the research and application of defect sites in single-chirality SWCNTs.

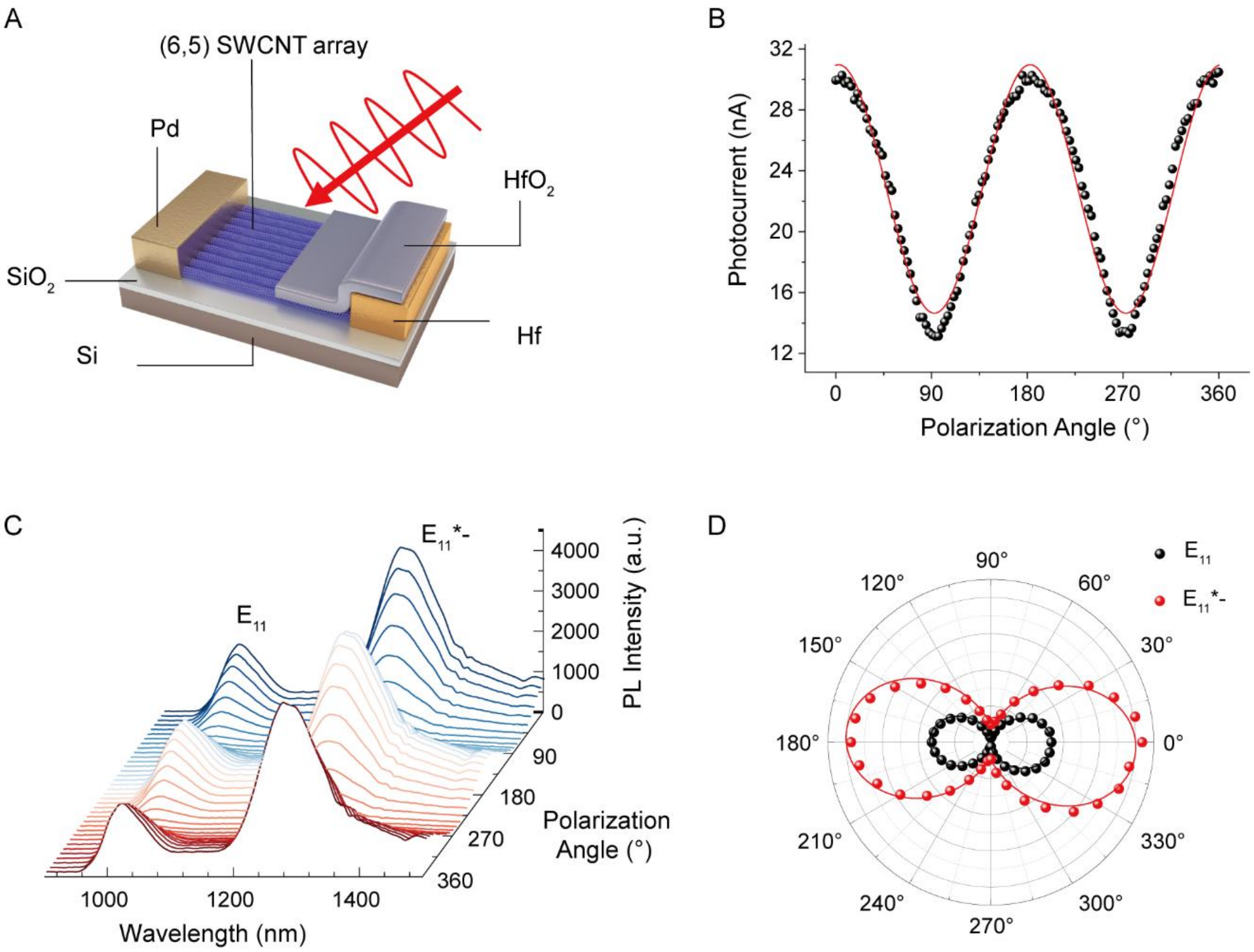

**Fig. 4. Anisotropic electro-optical and optical properties enabled by the aligned (6,5) SWCNTs.** (**A** and **B**) NIR light-detecting diodes based on aligned (6,5) SWCNTs. (A) Schematic diagram showing the (6,5) SWCNT–based NIR light-detecting diodes. (B) The photocurrent observed in the diodes when adjusting the polarizations of the incident light. The wavelength of the incident light is 1010 nm, which corresponds to the $E_{11}$ excitonic transition of (6,5) SWCNTs. (**C** and **D**) Angle-dependent PL of covalently-modified (6,5) SWCNT arrays. (C) Polarized PL spectra and (D) polar plot of the PL intensity of $E_{11}$ and $E_{11}^{*-}$ for different polarizations of the analyzer.

## Outlook

We demonstrate that the MISA method enables the preparation of high-density (more than 100 SWCNTs/μm) and well-aligned ($S_{2D}$ = 0.95) SWCNT arrays in a mono-layer from both organic and aqueous dispersions. Furthermore, we achieve the fabrication of high-density arrays of single-chirality and enantiomer-pure SWCNTs such as (−)(6,5) and (+)(6,5) SWCNTs, and explore their optical and optoelectronic properties, which have shown significantly anisotropic, exciton-related phenomena.

The MISA method enables fundamental studies on the physical properties of single-chirality and enantiomer-pure SWCNT arrays. First, enantiomer-pure arrays offer a platform to probe chiroptical effects and emergent properties. Second, in high-density arrays of single-chirality SWCNTs with precise pitch, there is periodicity perpendicular to the alignment direction, raising the question of whether such 1D systems exhibit collective quantum phenomena. Furthermore, the stacking of single-chirality arrays with distinct chiralities and twisted angles may give rise to new properties, representing a class of heterojunctions fundamentally different from those based on single 1D materials and 2D materials. Beyond SWCNTs, the high compatibility of the MISA strategy also suggests its potential for assembling other 1D nanomaterials.

**Acknowledgments:**

**Funding:**

National Natural Science Foundation of China (22120102004, U23A2085)

Ministry of Science and Technology of the People's Republic of China (2022YFA1203301)

Beijing National Laboratory for Molecular Sciences (BNLMS- CXTD-202001)

**Author contributions:**

Conceptualization: Yan Li

Methodology: Yuguang Chen, Yanzhao Liu, Zilong Qiu

Investigation: Yanzhao Liu, Zilong Qiu, Yuguang Chen

Funding acquisition: Yan Li

Supervision: Yan Li

Writing: Yanzhao Liu, Zilong Qiu, Yan Li

**Competing interests:** Yan Li, Yuguang Chen, and Zilong Qiu are inventors on a patent related to this work, granted to Peking University (Patent Number: ZL 2023 1 0648758.5). The other authors declare that they have no competing interests.

**Data and materials availability:** All data are available in the main text or the supplementary materials.

**Supplementary Materials**

Materials and Methods

Figs. S1 to S28